\documentclass[%
 reprint,
 superscriptaddress,
 amsmath,amssymb,
 aps,
 floatfix,
]{revtex4-2}
\usepackage{csquotes}
\usepackage{graphicx}
\usepackage{siunitx}
\usepackage{csquotes}
\usepackage{bm}
\usepackage{physics}
\usepackage{mathtools}
\usepackage{amsfonts}
\usepackage{amssymb}
\usepackage{amsmath}
\usepackage{diagbox}
\usepackage{hyperref}
\usepackage{booktabs}
\hypersetup{linktocpage,colorlinks,citecolor={blue},pdfdisplaydoctitle=true,pdfpagemode=UseOutlines,bookmarksnumbered=true}
\usepackage{dsfont}

\begin{document}
\title{A Christmas Story about Quantum Teleportation}
\author{Barry W. Fitzgerald}
\email{b.fitzgerald@tue.nl, barry.w.fitzgerald@gmail.com}
\affiliation{Eindhoven University of Technology, Eindhoven, Netherlands}
\affiliation{School of Biomolecular and Biomedical Science, University College Dublin, Dublin, Rep. of Ireland}
\author{Patrick Emonts}
\affiliation{Instituut-Lorentz, Universiteit Leiden, P.O. Box 9506, 2300 RA Leiden, The Netherlands}
\author{Jordi Tura}
\affiliation{Instituut-Lorentz, Universiteit Leiden, P.O. Box 9506, 2300 RA Leiden, The Netherlands}
\date{\today}

\begin{abstract}
Quantum teleportation is a concept that fascinates and confuses many people, in particular given that it combines quantum physics and the concept of teleportation.
With quantum teleportation likely to play a key role in several communication technologies and the quantum internet in the future, it is imperative to create learning tools and approaches that can accurately and effectively communicate the concept.
Recent research has indicated the importance of teachers enthusing students about the topic of quantum physics. 
Therefore, educators at both high school and early university level need to find engaging and perhaps unorthodox ways of teaching complex, yet interesting topics such as quantum teleportation. 
In this paper, we present a paradigm to teach the concept of quantum teleportation using the Christmas gift-bringer Santa Claus. 
Using the example of Santa Claus, we use an unusual context to explore the key aspects of quantum teleportation, and all without being overly abstract.
In addition, we outline a worksheet designed for use in the classroom setting which is based on common naive conceptions from quantum physics.
This worksheet will be evaluated as a classroom resource to teach quantum teleportation in a subsequent study.
\end{abstract}

\maketitle

\section{Introduction\label{sed:introduction}}

To many, the field of quantum physics is one of the most fascinating, yet perplexing fields of modern physics.
Despite its counter-intuitive character, quantum physics is predicted to have a significant impact on the technologies that we use for our future communication and online security~\cite{wehner_quantum_2018,wei_towards_2022,sridhar_review_2023}.

Nevertheless, when it comes to education on the topic of quantum mechanics, there are many obstacles. 
Students struggle with the abstractness of quantum physics and often rely on their experiences with classical physics to navigate the concepts from quantum physics~\cite{krijtenburg-lewerissa_insights_2017, bouchee_investigating_2023}. 
In addition, the language of quantum physics and its probabilistic aspect make learning difficult for students~\cite{bouchee_investigating_2023}.

Among the many concepts in quantum physics, one of the most intriguing is that of quantum teleportation. 
This process differs considerably from matter teleportation, a concept popularised in various fictional devices such as the transporter in Star Trek~\cite{krauss_physics_1995}.
Matter teleportation is associated with the transfer of people or objects from one point to another, which is impossible according to our current understanding of the laws of physics.

Quantum teleportation is possible though, and unlike fictional matter teleportation where atoms and molecules are moved from one location to another, quantum teleportation can be used distribute quantum information between two locations separated by distances of \SI{1000} {\km} or even further~\cite{yin_satellite-based_2017}.
In the future, quantum teleportation will play a key role in setting up quantum networks, a prerequisite for more intricate algorithms like quantum key distribution, a protocol at the heart of quantum-secure communication~\cite{bennett_quantum_2014,zapatero_advances_2023}. 

Therefore, it is imperative that the fundamentals, the possibilities, and the limitations of quantum teleportation are shared in a clear and novel manner with the next generation of quantum physicists at high schools~\cite{singh_precollege_students_qt_2022} and that educators have the required training to facilitate this \cite{sutrini_teacher_professional_development_2023}.
The second quantum revolution, which promises advancements in communication (based in part on quantum teleportation), sensors, computers, and simulations, "poses new challenges for the new emerging workforce who will develop these quantum technologies or work with them"~\cite{greinert_quantum_workforce_2023}.
But addressing the technological challenges is just part of the problem. There are numerous challenges with regards to teaching students in the first place, such as finding the academic talent to teach students and overcoming the lack of financial investment in key educational infrastructure and equipment~\cite{aiello_quantum_workforce_2021}.
In addition, there is an urgency for such teachings as quantum physics starts to become an even greater part of everyday lives, particularly in the face of pseudo-knowledge and misinformation~\cite{introne_how_2018}, which has also afflicted quantum science~\cite{meyer_mediahype_2023}. 

Let us not sugarcoat things though, quantum teleportation -- and quantum physics in general -- are difficult topics to grasp and to explain, thus creating issues for the educator and the student alike. 
Therefore -- in the high school classroom, powerful, innovative, and stimulating analogies are needed to scaffold the delivery of learning objectives associated with quantum physics and highlight the significance of quantum technologies for our future communications. 

Traditionally, when explaining communication between two parties or individuals in computer science and quantum physics, Alice and Bob are the characters of choice.
These fictional characters have been around since the late 1970s~\cite{rivest_method_1978}, and since then a collection of characters have been introduced to represent various participants in safe and malicious cryptographic communications. 
For instance, Eve or Yves is an eavesdropper on communications between Alice and Bob, while Wendy plays the role of a whistleblower. 

While Alice and Bob characters are part of the analogy of choice in quantum physics, when it comes to educating on quantum principles, an abstract communication protocol between two fictional characters may not be engaging enough for high school students. 
This presents an opportunity for an alternative and novel analogy to share the fundamentals of quantum teleportation. 

Here, we present a fun and somewhat unorthodox approach towards teaching quantum teleportation by drawing upon a character popular in many cultures -- the Christmas gift-bringer Santa Claus.
We use the process of delivering gifts around the world on Christmas Eve to explain the key concepts of quantum teleportation.
Thereafter, we present a worksheet for use in the classroom based on Santa Claus, and provide recommendations as to how this worksheet can be used in the classroom. 
The abstractness of quantum physics in general is known to hinder student learning of the topic~\cite{bouchee_towards_2022}. 
Therefore, the combination of the Santa Claus-based example and the worksheet are intended to provide a two-fold approach towards learning about quantum teleportation where the students first engage with the content, and then they learn through application and discussion.

\section{Pedagogical principles}
There are numerous previous studies dedicated to teaching quantum physics in general, the number of studies focused on quantum teleportation is limited.
Here, we consider recent works on developing approaches to teach quantum teleportation and the pedagogical approach used in both works.
We then outline the approach that we have adopted in our study before finally considering the importance of using such materials to enthuse both educators and students.

\subsection{Past studies on teaching quantum teleportation}
The first example is part of a larger research project with the aim of developing a 20-hour course for high school students on quantum computing where quantum teleportation is a sub-topic in the course~\cite{satanassi_quantumsecondary_2022}.
As part of their pilot study, the materials were evaluated with a small group of students between 16 and 17 years old.
The researchers set out to create resources on quantum teleportation and to find ways to make students aware of the revolution in quantum technologies. 

Design of the classroom materials on teleportation followed the model of education reconstruction (MER)~\cite{duit_mer_chaostheory_1997}, which consists of three main parts. 
First, the researchers catalogued existing teleportation experiments to find the one most suitable for use in the classroom. 
Second, student conceptions (naive or otherwise) in learning in general quantum physics were taken into account so that the content was made accessible to secondary students. The authors relied on their combined experience in the teaching of quantum physics.
Third, the materials were then tested in the classroom with a small group of students.
Via this process, the researchers in~\cite{satanassi_quantumsecondary_2022} developed a resource that followed the quantum teleportation experiment in Vienna of photons over a distance of 600 metres across the river Danube from 2004~\cite{ursin_danube_2004}.

In the second example, the researchers present a quantum curriculum transformation framework (QCTF), which can be used to create new materials for quantum technologies courses~\cite{goorney_qt_uni_2023}. In the paper, the researchers focus on examples pertaining to quantum teleportation. 
The materials in the paper are certainly suitable for higher education, but they could also be implemented with the same group of students from the previous study considered ~\cite{satanassi_quantumsecondary_2022}.

Central to the design of their resources are a number of components. 
First, the researchers used the European Competence Framework for Quantum Technologies, which is a tool designed by the EU's Quantum Flagship, as the basis for the materials to be created. 
Second, they identify three main skills to be targeted with the materials: namely theory and analytics, computation and simulation, and experiment and real world. 
The authors note that quantum teleportation, which encapsulates key concepts such as superposition and entanglement, can be used to develop all or some o these skills~\cite{goorney_qt_uni_2023}. 
Thereafter, the authors present descriptive (text), symbolic (mathematical) and graphical (pictures and diagrams) visual representations of the quantum teleportation process for the three main skills outlined above.
The final key aspect relates to the teaching approach for the materials. The authors argue that there are \enquote{infinitely possible ways} to teach quantum teleportation in the classroom when the requirements are for learners to understand the theory of quantum teleportation and recognise applications in the real world. 
Suggested teaching approaches include the open exploration of the materials by the student to guided learning where students are asked to focus on specific details~\cite{goorney_qt_uni_2023}.

\subsection{This study}
In this study, we align with the approach of Satanassi \textit{et al.}~\cite{satanassi_quantumsecondary_2022}, where the design process is inspired by two of the steps from the model of education reconstruction. 
First, rather than using an existing quantum teleportation, we devised a new explanation for quantum teleportation with a connection to popular
culture without too many technical details. 

Second, in creating the content and the worksheet, we took into account some naive conceptions of students in two different ways. 
We looked at a selection of the recent research on quantum physics to gain an appreciation for the concepts that may confuse students~\cite{krijtenburg-lewerissa_insights_2017, bouchee_investigating_2023, Bouchee_phdthesis_2023, bouchee_towards_2022, singh_precollege_students_qt_2022}.

\subsection{Importance of enthusing educators and students}
Finally in this section, we reflect on the importance of enthusing students about the concept of quantum teleportation, and quantum physics in general.
By means of example, in a recent PhD thesis, one researcher reflected on their own experiences in the high school classroom when teaching quantum physics and noted that they struggled to enthuse physics teaching about quantum physics~\cite{Bouchee_phdthesis_2023}.
Their experiences formed the basis of their research in which they explored the use of freely-available digital resources to teach high school students about quantum physics.

In one part of their study involving three high school physics teachers, the researcher noted that all three teachers struggled to enthuse their students~\cite{bouchee_investigating_2023}.
This then led to a compromised educational experience for teachers and students alike. 
In fact, two teachers and a number of the students in their respective classes all highlighted the need for practical and technological examples and applications of quantum physics~\cite{bouchee_investigating_2023}.
While a range of digital materials consisting of animations and interactive simulations~\cite{QuVis_website,PhET_website} are available to the educator, many of these rely on traditional contexts similar to those seen in textbooks.

The outcomes of the aforementioned study motivate in part the content and worksheet presented here.
In addition, one author has experience in the design and evaluation~\cite{fitzgerald_exploring_2018, fitzgerald_how_2020, plotz_superheroes_2021} of lesson materials based on popular culture (in this case for teaching the electromagnetic spectrum), which has proven to be effective at engaging with students. 
 
In summary, the content presented here is designed to enthuse educators and students alike in three ways. 
First, it presents a concept from quantum physics in an unconventional manner, using the gift-bringer Santa Claus. 
Second, it utilises a fictional character that would be recognisable in educational settings in many countries. 
Third, the example is based upon a practical implementation of the principles of quantum teleportation, albeit an unusual practical example, and does not focus solely on mathematical notation, as in previous classroom implementations~\cite{satanassi_quantumsecondary_2022}.

\section{Introduction to Quantum Teleportation\label{sec:background}}
Before presenting the exemplar for quantum teleportation based on Santa Claus delivering gifts on Christmas Eve, we first outline the key aspects of quantum teleportation.
By definition, quantum teleportation is a protocol to transfer quantum information from one location to another.
For this definition, several questions in relation to quantum teleportation need to be addressed such as \enquote{What is a protocol?} and \enquote{What is quantum information?}.
We now address these by making reference to the paradigm of a quantum state transfer via quantum teleportation between the traditional characters of Alice and Bob.  

\subsection{What is a protocol?\label{sec:sub_protocol}}
Let us imagine that Alice needs to tell Bob something very important, but they are not in the same room. 
A protocol describes the steps needed by Alice and Bob to successfully transfer information.
A simple example of a protocol is a phone call between Alice and Bob. 
If Alice wants to speak to Bob on the phone, she must dial Bob's phone number, and then press the call button.
On Bob's side, he has to accept the call so that Alice can speak to him and share the information.
The protocol relies on both parties having certain resources (a telephone each and Alice needs Bob's number) and performing well-defined actions (Bob must accept Alice's call before hearing what Alice has to say).

\subsection{Quantum information and quantum states\label{sec:sub_quantum}}
In quantum teleportation, the protocol is similar to a phone call in the sense that information is transferred, but it is quantum information that is transferred rather than spoken information. 
In the classical computers that you will find in almost every electronic device on the planet from smartphones to flying drones, information is encoded in bits as $1$s and $0$s, the basic units of classical information in computers.
The bits $0$ and $1$ can represent everything from a traffic light being red or green (two states) to a coin showing head or tails.
In real devices, these $1$s and $0$s can be, for example, realized by the orientation of tiny magnets in a hard drive device where the magnets can be orientated in one of two directions.

On the other hand, quantum information is information stored in \textit{qubits}. 
Short for quantum bit, a qubit is the fundamental unit of information in quantum computing. 
Like the bit in a classical computer, a qubit also has two states, and while these states are not $1$ and $0$, there are similarities.  
Rather than using $1$ and $0$, the qubit states are $\ket{1}$ and $\ket{0}$.
The strange notation $\ket{\cdot}$ indicates that qubits are different from classical bits.
The key difference between classical and qubits is that qubits can be in a quantum superposition.

\subsection{What is quantum superposition?\label{sec:sub_superposition}}
\begin{figure}
    \centering
    \includegraphics[width=0.8\columnwidth]{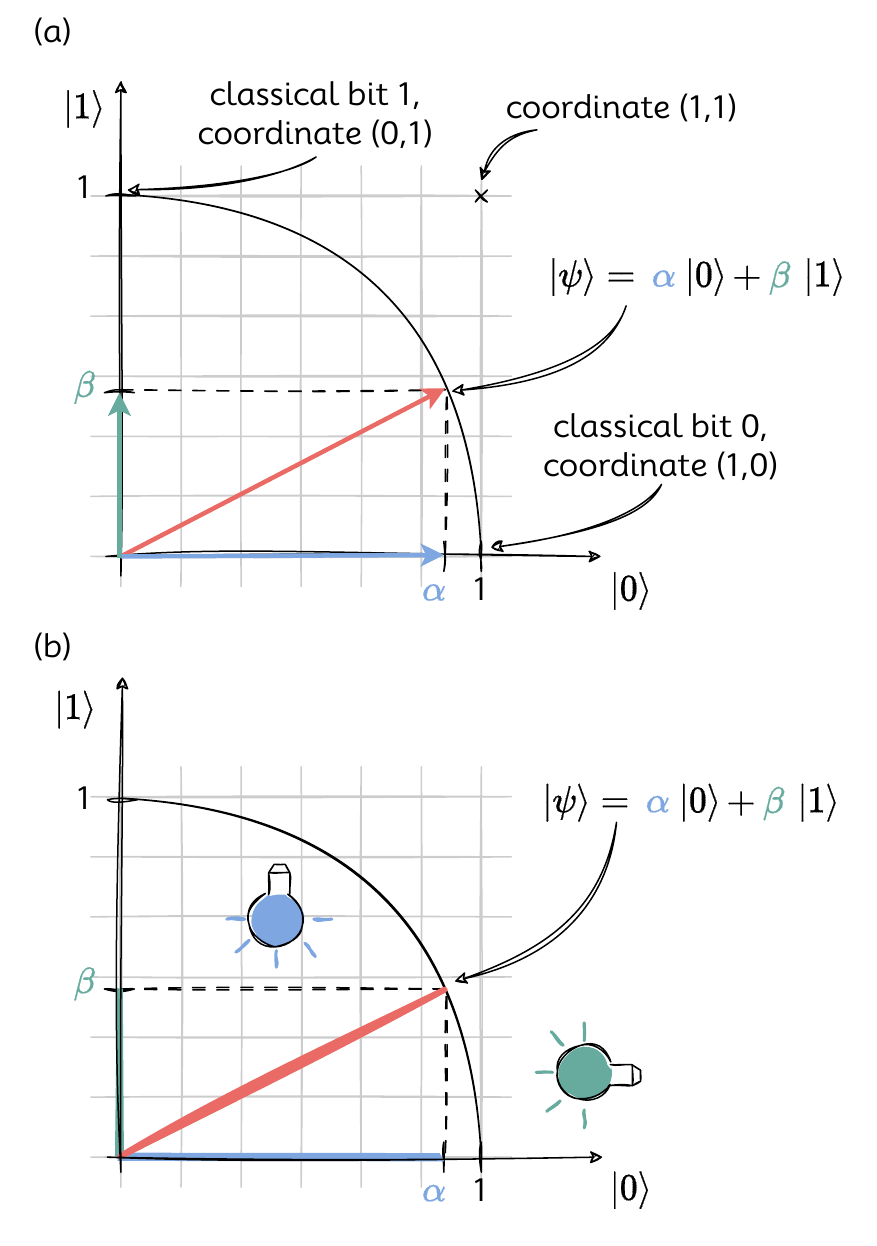}
    \caption{Illustration of a superposition of states. 
    \emph{(a)} Qubits can be illustrated as directions in a two dimensional coordinate systems.
    The vectors $\ket{0}$ and $\ket{1}$ take on the roles of the $x$- and $y$-axis, respectively.
    Any vector that does not point in the direction of one of the axis, is a superposition.
    \emph{(b)} Measuring a quantum state can be visualized as a stick pointing in the direction of the quantum states and the shadows cast by two lamps.
    The longer the shadow, the higher the probability to measure that particular outcome.
    In the case shown here, the shadow along the $\ket{0}$ axis is longer, so this has a higher probability of being measured. 
    }
    \label{fig:superposition}
\end{figure}

So, what is the difference then between bits in classical computers and qubits in quantum computers?
An easy way to realize that qubits can represent more information than classical bits is by visualizing the qubits in a coordinate system.
In Figure~\ref{fig:superposition}(a), the two quantum bits are used as the axes of a coordinate system.
Classical bits $0$ and $1$ are always located exactly on these axes: there are only two choices $0$ and $1$.
However, there is a lot of white space left between the two axes, where classical bits can never be.

This same is not true for qubits.
The marvels of quantum physics means that a qubit can be in any state between the two axis.
This is known as quantum superposition. 
We can actually understand the word superposition quite literally.
The states $\ket{0}$ and $\ket{1}$ form the axis of the coordinate system.
Instead of the $x$- and $y$-direction, the directions of the coordinate system are called $\ket{0}$ and $\ket{1}$.
These vectors actually represent the same vectors as in everyday vector calculus
\begin{align}
    \begin{aligned}
        \ket{0}&=\mqty(1\\0) \qq{unit vector in $x$ direction,} \\
        \ket{1}&=\mqty(0\\1) \qq{unit vector in $y$ direction.}
        \label{eq:qubit_computational_basis}
    \end{aligned}
\end{align}

By super-imposing these two directions with certain coefficients or weights $\alpha$ and $\beta$, we can form an arbitrary quantum state, which we write as
\begin{align}
    \ket{\psi}=\alpha \ket{0} + \beta \ket{1}.
    \label{eq:qubit_superposition}
\end{align}
Here, $\ket{\psi}$ represents the superposition state of the qubit. 
Any direction in the two-dimensional plane [cf. Fig.~\ref{fig:superposition}(a)] can be understood as a quantum state.
Since only the direction encodes a quantum state, the length of the vector does not play a role.
This means that we can limit ourselves to vectors of length one.
In the illustration, this is illustrated by the black arc connecting the $\ket{0}$ and the $\ket{1}$ axis.
If we translate the condition to the coefficients of the state, we notice that they must fulfill $|\alpha|^2+|\beta|^2=1$.

To be fair, the coefficients $\alpha$ and $\beta$ could be taken to be complex numbers, so instead of a circle in the plane, the actual mathematical model of the qubit is a sphere, but that is a technical issue we do not need to consider here.
Therefore, we use a simplified model where $\alpha$ and $\beta$ are real numbers and therefore can be interpreted as the $x$- and $y$- coordinates of a circle, respectively.

Although we can prepare an arbitrary quantum state, observing which quantum state we have is more difficult.
Performing a measurement on a qubit means that this qubit will return to $\ket{0}$ or $\ket{1}$. Why is that so? In fact, it is important to note that we have to display the result of that measurement on a classical display (there are no such things as displays for quantum information).
Thus, we must either get $0$ or $1$, corresponding to $\ket{0}$ and $\ket{1}$, respectively.
If we look at Figure~\ref{fig:superposition}, we know two notable points already: The coordinate axes corresponding to the classical bits $0$ and $1$.

So, what happens if the qubit is neither exactly $\ket{0}$ nor $\ket{1}$? 
How does it decide which one to become after the measurement? 
The answer is related to $\alpha$ and $\beta$.
If we measure a quantum state, the length of $\alpha$ and $\beta$ are related to the probabilities of measuring the state in $0$ and $1$.
In technical terms, the probabilities are given by the squared projections of the vector to the axis $\ket{0}$ and $\ket{1}$.

Intuitively, the projection process can be imagined as shadows cast by the vectors along the two axes as shown in Fig.~\ref{fig:superposition}(b).
If we think of the quantum state as a stick, the projections on the axes $\ket{0}$ and $\ket{1}$ are the shadows cast by the green and the blue lamp.
The green lamp casts a shadow onto the $\ket{1}$ axis.
The length of that shadow is equivalent to the amount of the stick that points in the $\ket{1}$ direction.
In other words, it is the part of the quantum state in the $\ket{1}$ direction in Eq.~\eqref{eq:qubit_superposition}.
However, we cannot observe the coefficients $\alpha$ and $\beta$ directly, unless we have many copies of the quantum state.
If you want to measure them, you have to prepare and measure the quantum state many times.
The number of times that either $0$ or $1$ is measured can be used to estimate the values for $\alpha$ and $\beta$.

Importantly, a measurement changes the quantum state. 
Remember, it is not really a stick after all.
After measurement, the state will be pointing exactly in the direction that was measured (and its length will be one again).
The quantum superposition is lost, and the qubit points either in the $\ket{0}$ or $\ket{1}$ direction, according to the measurement result.
The information that we get as the result of a measurement is classical again: it points exactly along one axis.
In quantum physics, this is called the collapse of a quantum state.

The collapse of the quantum states leads to an important feature of quantum information.
A general quantum state can never be copied or cloned, it can only be transferred~\cite{park_concept_1970, wootters_single_1982, dieks_communication_1982}.
This is in stark contrast to our experience with classical bits.
Every film that you stream is copied from a server and sent to you.
This is not possible with quantum states, because trying to learn the exact values of $\alpha$ and $\beta$ destroys the quantum state itself.

But what do qubits look like in the real world?
In analogy to the variety of systems where we can store classical information (traffic lights, hard drives, paper, etc.) there is a whole plethora of platforms that can encode quantum information.
For example, qubits in quantum devices can be represented by the spin of an electron (an electron has two spin values -- spin up and spin down) or the polarization of photons where the two states are vertical polarization or horizontal polarization, just to name two.
Here, we do not focus on the technology used to store the qubits, and instead focus on how qubits relate to quantum teleportation.

\subsection{What is quantum entanglement?\label{sec:sub_entanglement}}
In the macro-world that we live in, we can experience entanglement on a daily basis. 
Just think of entangled shoelaces, ropes, or even long hair. 
However, in the quantum world of qubits, quantum entanglement is something different. 
It is a special kind of interrelationship that two or more quantum systems (qubits in our example) have with each other. 

Let us go back to the example of Alice and Bob from the description of a protocol.
Imagine that Alice and Bob have one qubit each, $q_A$ and $q_B$, respectively.
Let us check which quantum states we can realize with two qubits.
Using the intuition from the one-qubit case, we can list all directions the qubit could point to.
In the one-qubit case, these are $\ket{0}$ and $\ket{1}$, which is just a list of all classical values a bit can take ($0$ and $1$).
Let us list the classical states two bits can take, and there are four values in total -- $00$, $01$, $10$, and $11$.
Thus, two qubits can, in general, take all states of the form $\ket{\psi}=\alpha_1\ket{0_A,0_B}+\alpha_2\ket{0_A,1_B}+\alpha_3\ket{1_A,0_B}+\alpha_4\ket{1_A,1_B}$, where the $\alpha_i$'s determine the superposition of the four possible states. 
As in the one-qubit case, we just enter the numbers in fancy brackets to show that we are not considering classical bits anymore.
The subscripts $A$ and $B$ in the formula are inserted to remind us that one qubit belongs to Alice and one qubit belongs to Bob.
The mathematical details of this notation are explained in Appendix~\ref{app:teleportation}.

Some states with two qubits are special, since we can learn something about one qubit by measuring the other qubit.
These states are said to be entangled.
One special case is the maximally entangled state
\begin{equation}
    \ket{\phi}=\frac{1}{\sqrt{2}}(\ket{0_A,0_B}+\ket{1_A,1_B}).
    \label{eq:maximally_entangled}
\end{equation}
This state is a superposition state of the two-qubit states $\ket{0_A,0_B}$ and $\ket{1_A,1_B}$.
As in the one qubit case, we can only measure the outcomes that are in the superposition.
The state will always collapse to either the classical state $00$ or $11$.
The probabilities are determined by the coefficients in front of the qubits.

Let us consider what happens if we measure the maximally entangled state.
If Alice measures her state, she obtains either $0$ or $1$, both with an equal \SI{50}{\percent} probability, since the coefficients in front of both qubits are the same.
We cannot predict which result she will get, as it is probabilistic after all.
Once she has measured the status of her qubit, Alice will know Bob's result! 
It has to be the same as her own result.
First, if Alice measures $0$, then the state collapses to $\ket{0_A,0_B}$ and, as a result, Bob will also measure $0$.
Second, if Alice measures $1$, then the basis state collapses to $\ket{1_A,1_B}$ and, as a result, Bob will also measure $1$.

However, not all two-qubit states are entangled.
It is important to distinguish between quantum correlation and classical correlation.
The simple fact that Alice and Bob measure the same outcome is not surprising.
This effect can be easily obtained with classical objects.
For instance, imagine you can pick either red or blue socks in the morning.
Since you like symmetry, you always choose both socks to have the same color.
Thus, if someone sees only one of your socks, that person can be certain that the other sock has the same color.
So far, this is the same as Alice measuring her qubit and knowing what Bob will get.
Nonetheless, the randomness of the outcome in the case of qubits is a critical part of the story.
Alice's outcome is only fixed once she measures it, and Bob will still have the same outcome no matter if Alice measures $0$ or $1$.
This is the beauty and puzzle of quantum correlations.

We note that we said nothing about the distance between Alice and Bob so far.
They could even be on the opposite sides of the galaxy and the argument would still work.
The fact that Alice and Bob get the same result (either both of them get $0$ or both of them get $1$) does not require that they are close to each other.
Would it then be possible to communicate instantaneously across large distances with some sort of quantum telephone? Unfortunately, no.
This would violate one of the key postulates of special relativity: no information transfer can happen faster than the speed of light.

Unfortunately, we cannot build a quantum telephone that operates in this way.
In the case above, Alice knows Bob's measurement, but Bob does not know Alice's measurement. 
Bob would have to check his qubit himself, and because the qubits are maximally entangled, he would then know the value of Alice's qubit. 
If Alice wanted to tell Bob about her measurement directly, she would have to somehow send that information to Bob in the form of a standard transmission such as a text file or short audio message.
This latter communication is limited by the speed of light.

\subsection{What is entanglement distribution?\label{sec:sub_entanglement_distribution}}
To use entanglement between qubits for quantum teleportation, we have to find a method to distribute entanglement between Alice and Bob.
As the aim is to transmit quantum information from Alice to Bob over a distance, it would be counterproductive if they had to meet in person to exchange entangled qubits.

The solution to distributing entanglement can come in the form of entangled photons. 
As photons travel at the speed of light, this makes them perfectly suited to be ``entanglement messengers''. 
There are a variety of quantum optics experimental approaches available that can produce entanglement between two photons.
We can send them to Alice and Bob to have them share an entangled pair.
Once the entangled photons arrive at Alice and Bob, they must store the quantum state in some quantum device since photons cannot be stopped and always travel at the speed of light.
Such a storage process has been demonstrated, but is still under active development~\cite{sangouard_quantum_2011,munro_inside_2015,langenfeld_quantum_2021}.

\subsection{All leading to quantum teleportation}
So far, we have covered the fundamentals of quantum information and entanglement. 
Here, we show how all of this comes together to facilitate quantum teleportation.

For the protocol to be successful, three qubits are required. 
Alice is assigned two qubits $\ket{\psi_{A_1}}$ and $\ket{\psi_{A_2}}$, while one qubit $\ket{\psi_B}$ is intended for Bob's side of the communication. 
Our goal is to transfer the quantum information from Alice's qubit $\ket{\psi_{A_1}}$ to Bob's qubit $\ket{\psi_B}$.

\begin{figure}
    \centering
    \includegraphics[width=\columnwidth]{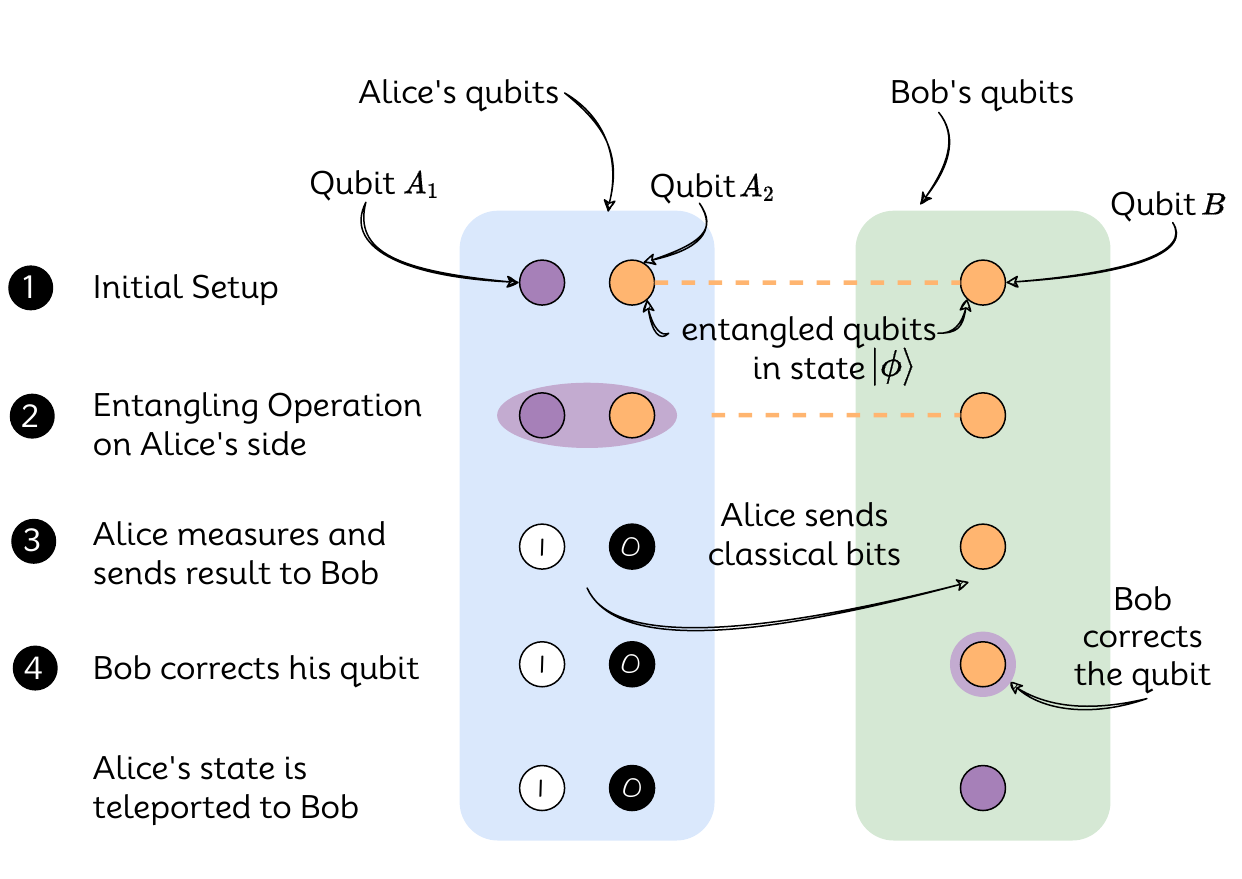}
    \caption{Illustration of the quantum teleportation protocol.
        The protocol consists of four steps.
        In the first step, the entangled qubit is shared between Alice and Bob.
        With local operations, Alice entangles the target qubit with the entangled pair (step 2).
        Then, Alice measures both of her qubits (step 3) and tells Bob the result of the measurement via classical communication.
        He can then correct his qubit (step 4).
    }
    \label{fig:teleportation}
\end{figure}

Alice's second qubit $A_2$ is maximally entangled with Bob's qubit $B$ leading to
\begin{align}
    \ket{\phi_{A_2B}}=\frac{1}{\sqrt{2}}\left( \ket{0_{A_2}0_B} + \ket{1_{A_2}1_B}\right).
\end{align}
Note that this is the same as Equation~\eqref{eq:maximally_entangled}, just with adapted indices.
Although the notation seems a bit cumbersome, we use the indices here to highlight who holds which qubit.
Now, the state of the whole system comprises qubit $\ket{\psi_{A_1}}$ and the entangled pair $\ket{\phi_{A_2B}}$, which is distributed across Alice and Bob, and can be written as is $\ket{\psi_{A_1} \phi_{A_2B}}$.
The overall goal is to transfer the state of Alice's first qubit $\ket{\psi_{A_1}}$ to Bob's qubit $\ket{\psi_B}$.
The initial setup is illustrated in step 1 of the quantum teleportation protocol shown in Figure~\ref{fig:teleportation}.

The second step is to share the entanglement of Alice's second qubit (denoted $A_2$) and Bob's qubit with Alice's first qubit $A_1$ as illustrated in step 2 in Figure~\ref{fig:teleportation}.
To achieve this, Alice performs two operations on her side (cf. Appendix~\ref{app:teleportation} for further details).
This manipulation on Alice's side leads to a special structure of the overall state of the system whereby Bob's qubit is in one of four states.
Each of these states only slightly differs from the target state $\ket{\psi_{A_1}}$.
Through the application of some correction operations, Bob could obtain the target state $\ket{\psi_{A_1}}$, but he does not know which operation he should perform as of yet.

In step 3, Alice obtains the information necessary for Bob's correction.
Alice measures her two qubits and the superposition between the qubits of Alice and Bob collapses.
As a result, Alice's pair of qubits $\ket{\psi_{A_1}}$ and $\ket{\psi_{A_2}}$ are then in one of four output pairs, which are $00$, $01$, $10$, or $11$.
Now, Alice calls Bob and tells him the result of her measurement.

Knowing the result on Alice's side, Bob can correct his qubit (step 4).
After the correction, Bob's qubit $\ket{\psi_B}$ then contains the state that was originally stored in Alice's qubit $\ket{\psi_{A_1}}$.

\section{Santa Claus and quantum teleportation\label{sec:santa_example}}
The fundamentals of quantum teleportation were presented in the previous section, and we appreciate that this might have led to some \enquote{quantum overload} in terms of learning about quantum teleportation.
Although we used the traditional exemplar of state transfer between Alice and Bob to explain the various components of quantum teleportation, for a high school student this could (and should) be presented in a more engaging fashion.

Therefore, to invigorate the learning of quantum teleportation, we outline here a detailed example motivated by the Christmas deliveries of the famous gift-bringer Santa Claus.
On Christmas Eve, Santa Claus sets forth from his workshops in the Arctic Circle to deliver presents to expectant children around the world. 
In logistical terms, this is one of the most arduous delivery schedules undertaken by anyone in the history of mankind. 

The adventures of Santa Claus at Christmas are a key part of many cultures globally, with millions of young children asking for presents from Santa Claus each year. 
While Santa Claus is not the primary gift-bringer in all countries, many people around the world are familiar with the activities of Santa Claus at Christmas from popular culture content such as films and TV series.
In other words, we are using it as an accessible and fun paradigm for students.

We have selected Santa Claus for this exemplar as it is an unconventional context and Christmas presents are an interesting application of the quantum teleportation process.
While there are plenty of website articles and videos in relation to the potential science behind Santa Claus' delivery approach, there are few literature examples regarding the use of Santa Claus to communicate on science and engineering~\cite{fitzgerald_secret_2016, highfield_physicschristmas_2008}.

The narrative connecting Santa Claus and quantum teleportation presented here is used for illustrative purposes and to present the complexities of quantum teleportation in a new light.
Importantly, in our example we use the presents to represent the sharing of information. 
We are not suggesting in any way that this example infers that matter teleportation is possible. 

In the following, we tell the story and make the connection to section~\ref{sec:background} at the same time.
The full story is given in Appendix~\ref{app:story}.
\begin{figure*}
\centering
\includegraphics[width=\textwidth]{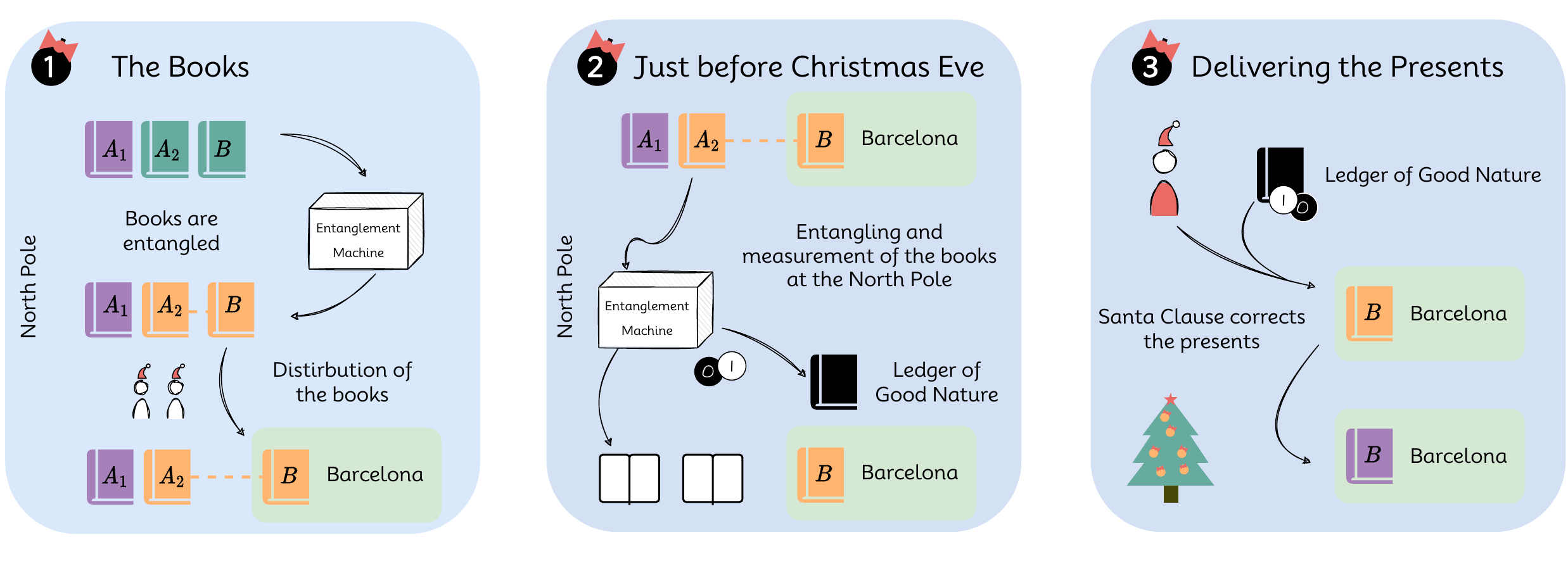}
\caption{Illustration of the quantum teleportation protocol as a Christmas story. 
Panel (1) shows the preparation of entangled qubits (books). The purple book contains the information to be teleported.
The connected, orange books are entangled.
In Panel (2) the actual teleportation protocol is executed.
The measurement result which is needed for the correction is recorded.
In Panel (3) Santa Claus corrects the book's content to ensure equality of the purple book's content in Panel (1) and the purple book in Panel (3).
}
\label{fig:process_santa_claus}
\end{figure*}

\subsubsection{Part 1: The Books}

It is Christmas, and this year, Santa Claus will be delivering one type of present to every child around the world - a book. 

During the protocol, the books represent the qubits while the content of the books represent the quantum information stored in the books.
Importantly, a book is never teleported, it is always an individual book's contents that is transferred.
The protocol involving Alice and Bob uses qubits which can be described with two basis states.
These basis states are represented by a blank page ($\ket{0}$) or a written page ($\ket{1}$).
A book contains much more information than just one qubit or one bit.
For the sake of the metaphor, we will pretend that the book's contents can be described by just one qubit.

Unfortunately, hiding presents in all homes across the world is a time-consuming and arduous task, which essentially takes the whole year and cannot be completed in a single night.
Santa likes to plan ahead, and he would love to distribute the books well ahead of time.
However, he will only receive the children's wishlist at the start of December.
Santa and his elves, therefore, resort to quantum teleportation to deliver the books on time for Christmas Eve.

To teleport the contents in the book to every child anywhere around the world, three books are required and they correspond to the three qubits needed for quantum teleportation. 
Two books $\ket{\psi_{A_1}}$ and $\ket{\psi_{A_2}}$ are based at the North Pole ($A$ stands for Arctic Circle), while the third book $\ket{\psi_B}$ is one that needs to be placed at the home of the child prior to Christmas Eve.
Here, we choose a child in Barcelona, hence the name $\ket{\psi_B}$.
Long before Christmas, all books are at one of Santa's workshops close to the North Pole.

First, before any present book is delivered to a home, Santa and the elves use a special Christmas entanglement machine to entangle the present book $\ket{\psi_B}$ with the workshop book $\ket{\psi_{A_2}}$.
The books are placed side-by-side in the machine, and the machine then evaluates if the content on any given page in both books will be either blank or full of text. 
Note, the entanglement machine cannot tell what the actual content is and nobody is allowed to open the books afterwards until the protocol is completed.

There are two key features here: the question that creates entanglement and the inability of the machine to tell the precise contents.
The machine asks the question whether a certain page in both books is either blank or has text.
Here, the two states of the page represent the two states of a qubit: $\ket{0}$ means blank and $\ket{1}$ stands for text.
Since the questions allows only two outcomes, both containing text or both blank, we ensure that the two-book system (two-qubit system) is in a state of the form $\ket{BB}+\ket{TT}$, where $T$ stands for text and $B$ stands for blank.
This is the same structure as the maximally entangled state in Equation~\eqref{eq:maximally_entangled}.
The two books are now playing the role of the maximally entangled state in the quantum teleportation protocol.

After the entangling procedure, nobody is allowed to open the book and the machine is not allowed to \enquote{read} the text on a given page.
This would correspond to measuring and the superposition created in the step would vanish.
We would be left with purely classical and random information, and not with a quantum state.

The elves' job is not done after entangling the books.
One half of the entangled book-pair has to find its way to the children's homes.
Santa's elves are expert at hiding the present books in homes, and do so in a top-secret manner.
This arduous task is completed over the course of the year so that the elves have time to visit all homes around the world. 
They have a lot of time since the actual present, which is the contents of the book stored in $\ket{\psi_{A_1}}$, will only be teleported on Christmas Eve.

Of course, it can happen that a child finds the present book at home, which would ruin their Christmas present from Santa Claus. 
This is not an issue though because if a child were to open the present book without proper completion of the quantum teleportation process, the entangled state would be broken. 
As a result, the child would see gibberish on the pages and it would make no sense.

The idea of opening the book early corresponds to measuring the quantum state.
The entanglement will be destroyed and the child is just left with random information that is unintelligible and does not make for a nice present for the child. 

\subsubsection{Part 2: Just before Christmas Eve}
The next stage of the process begins just before Santa Claus leaves his workshop on Christmas Eve.

First, for each child, the elves entangle the present book $\ket{\psi_{A_1}}$ with entangled book at the workshop $\ket{\psi_{A_2}}$ using the special Christmas entanglement machine. 
As before, the books are placed side-by-side in the entanglement machine, and the machine assesses whether on any given page from both books, there will be either a page full of text or a blank page in the two books. 
Then the elves add that information about the entanglement to a ledger for Santa Claus known as the \textit{Ledger of Good Nature} to keep a record of the entanglement process. 
Whether a child receives a book on Christmas or not is influenced by their level of good nature over the course of the year. 

The entanglement operation between the two books at the Arctic Circle workshop corresponds to the entanglement operation on Alice's side in the second step of the protocol [cf. Fig.~\ref{fig:teleportation}].
In this Santa Claus example, the second step and the third step of the protocol are combined.
The elves have to record the outcome of the measurement (step 3) because Santa needs the information to correct the book at the child's home in the final step of the process.
If Santa decides not to bring a gift to the child, he can do so at the last minute, simply by not correcting the book on the child's side.

As a result of the creation of entanglement between $\ket{\psi_{A_1}}$ and $\ket{\psi_{A_2}}$, the elves have also created entanglement between $\ket{\psi_{A_2}}$ and $\ket{\psi_B}$ from Part 1 of the process. 

The reason that book $\ket{\psi_{A_2}}$ is used in this process, and not $\ket{\psi_{A_1}}$ and $\ket{\psi_B}$ right away, is that the books stored in $\ket{\psi_{A_1}}$ are not assigned to children until very close to Christmas. 
This is because the wishlists from children only start to arrive from the start of December approximately. 
Using the first entanglement pair allows the elves and Santa to prepare an entangled book well in advance of Christmas, and then once the book $\ket{\psi_{A_1}}$ is selected for the child, it is entangled with $\ket{\psi_{A_2}}$, and as a result with $\ket{\psi_B}$. 

The pair of entangled books is a tool to enable quantum teleportation later at a flexible and more opportune moment.
For instance, imagine bringing a fork and a plate to a buffet dinner, but now knowing what you are going to eat at the buffet dinner. 
All you know for certain is that you will use the fork and plate to eat some food, but you do not know yet what food you will be eating. 
It is the same process with regards to the use of the first pair of entangled books - the elves and Santa Claus know they are going to need them, but they do not know yet what they will send with them.

\subsubsection{Part 3: Delivering the presents}
On Christmas Eve, Santa Claus travels around the world to deliver the presents to the children. 
In effect, he travels around the world to complete the quantum teleportation process. 

During the journey, Santa Claus checks the \textit{Ledger of Good Nature} to see whether the child will receive a present or not.
If the child has been well behaved during the year, then the child will get the book that they have asked for. 
However, if the ledger indicates that the child has misbehaved too much, the child will not get the book that they asked for. 
Instead, they will get random, gibberish text in their book, because Santa does not correct the information in the book.

This part of the process is all about Santa Claus initiating the final step of the quantum teleportation process, the correction step.
The result of the ensuing measurement was recorded in the ledger.
Recall that $\ket{\psi_{A_1}}$ and $\ket{\psi_B}$ are linked to each other by the process. 

With this in mind, Santa Claus must complete the final step of the quantum teleportation by travelling around the world. 
When Santa Claus arrives at the home of each child, he quickly locates the hidden book $\ket{\psi_B}$, and corrects its contents according to the information in his ledger.
He sets up his present scanner with the right information from the \textit{Ledger of Good Nature}, which adapts the book's contents to ensure that the present book reads the same as the originally intended book stored in $\ket{\psi_{A_1}}$.

On Christmas morning, when the child opens the book $\ket{\psi_B}$, they can open the book without worrying about seeing gibberish as a result of breaking the quantum teleportation. 
Of course, if the child has been deemed to have misbehaved over the year, they will receive a book filled with gibberish text anyway!

\section{Classroom worksheet on Santa Claus and quantum teleportation\label{sec:santa_worksheet}}

In the previous section, we considered the process by which Santa Claus could deliver presents in the form of books around the world on Christmas Eve using the principles of quantum teleportation.
The key aspects of quantum teleportation are included using this fun and unconventional example, but it is also important to present students with the opportunity to learn through practice from this example with suitable support materials. 
In a recent study on the experiences of teachers and students in quantum physics lessons in the Netherlands, some teachers and students highlighted the lack of such materials for practice, while some teachers pointed out a lack of clear applications and context~\cite{bouchee_investigating_2023}.

Therefore, as a next step in our proposed quantum teleportation-Santa Claus lesson, we present a classroom worksheet [cf. Appendix~\ref{app:worksheet}] based on the Santa Claus exemplar that addresses both application and context, while also ensuring that the lesson is light on mathematical notation.  
In this case, the application is the distribution of quantum information - in the form of gifts - around the world by Santa Claus on Christmas Eve. 

\subsubsection{Worksheet design and student learning objective}
We have designed a worksheet similar in form to those worksheets that one author has used to support lessons on the electromagnetic spectrum using the superhero genre~\cite{fitzgerald_exploring_2018, fitzgerald_how_2020, plotz_superheroes_2021}, which were well received by students as demonstrated by results from a post-test survey of students.
We envision that the worksheet presented here can also have a similar impact on the teaching of quantum teleportation in the classroom, which we intend to evaluate as part of a future study. The goal of a future study would be to assess the content and worksheet suitability to communicating on quantum teleportation and use student input on the worksheet to revise the worksheet where required.

The worksheet has one primary goal, which relates to the student learning objective, and that is to give students the opportunity to demonstrate their understanding of quantum teleportation. 
This is achieved in two ways. First, the students are asked to define the key attributes of quantum teleportation such as a qubit, superposition, and the key steps in quantum teleportation. 
Second, with the final question, the students are invited to critically assess the quantum teleportation process and to identify the steps that require the most attention in the Santa Claus example. In terms of the revised Bloom's taxonomy~\cite{Krathwohl_revised_blooms_2002}, this question engages the higher levels of cognitive complexity such as analyse, evaluate, and create. 
As a secondary goal, we aspire to excite students about the real possibilities of quantum teleportation.

The worksheet questions are intended to address naive conceptions that students have in relation to quantum physics. Below we list such naive conceptions and the questions from the worksheet in Appendix~\ref{app:worksheet} that are intended to help address them:
\begin{itemize}

    \item Students struggle to transit from the deterministic world of their common experiences to the probabilistic world of quantum physics~\cite{bouchee_towards_2022} - Question 2;
    
    \item Students have difficulties with interpreting and comprehending counter-intuitive quantum physics concepts such as qubits and superposition~\cite{bouchee_towards_2022, greinert_quantum_workforce_2023} - Questions 1 and 2;

    \item Students struggle with the language of quantum concepts~\cite{bouchee_towards_2022} - Questions 4 and 5;
    
\end{itemize}
The only question not categorised here is question 3 from the worksheet, which simply asks the students to give the number of qubits needed for quantum teleportation. This question acts as the bridge between the first two questions where the students are asked to expand on two quantum concepts, and the final two questions where the students are required to give details related to a quantum teleportation process.

\subsubsection{Worksheet overview}
The quantum teleportation worksheet (cf. Appendix~\ref{app:worksheet}) is split into two sections. 
First, a motivation for the quantum teleportation worksheet based on the Christmas Eve adventures of Santa Claus is presented as well as the delivery expectations on Christmas Eve. 

Second, the students are then invited to put their own stamp on the quantum teleportation process. 
They are invited to select a present type that they would like to distribute on Christmas Eve. 
Thereafter, the students can confine the delivering process to a particular city, country, or continent.  
The only invariable part of the worksheet is the inclusion of Santa Claus. 

Finally, the students are posed a series of questions in relation to key steps of the quantum teleportation process as well as open questions to discuss with each other in relation to the process. 

This worksheet can be used to encourage students to immediately apply their new knowledge related to quantum teleportation. 
Additionally, the questions encourage critically thinking about the concept.
The combination of the worksheet and the lesson is intended to present a clear application of quantum teleportation, albeit using an unconventional and unlikely situation. 

Such an unusual approach towards covering quantum concepts in the classroom could be adapted for other learning objectives in quantum physics.
In a recent study exploring both teacher and student experiences of quantum physics lessons, students noted that the quantum physics topics are difficult, while teachers found it difficult to identify clear applications, which can lead students to question its relevance~\cite{bouchee_investigating_2023}. 
Furthermore, students have difficulty relating concepts from quantum physics to the macro-world in which they live~\cite{krijtenburg-lewerissa_insights_2017}.
Therefore, it is imperative that teachers are challenged to adopt new ways of enthusing students in their learning of quantum physics. 
An unconventional approach based on Santa Claus as outlined here with regards to quantum teleportation would certainly garner student attention given that it is based upon a popular culture icon. 
We envisage that a similar lesson plan and worksheet could be formulated for other concepts from quantum physics, although this is outside the scope of the current work.

\section{Conclusions}
In quantum physics, the concept of quantum teleportation is both fascinating and complex. 
As the field of quantum physics matures and the societal applications become a reality, quantum teleportation is a technique that looks destined to impact how we share quantum information over long distances. 
Despite its importance, the technique remains abstract and confusing to most working outside the field, in part due to a lack of relatable and engaging paradigms to explain the intricacies of the technique.

Therefore, we have presented an unusual context based on Santa Claus that covers the fundamentals of quantum teleportation. 
First, we have presented a lesson outline that incorporates the key steps of quantum teleportation as pertaining to the delivery of books by Santa Claus.
Second, we have designed a classroom worksheet to support this paradigm that can be used by the students to apply what they have learned in relation to quantum teleportation. 

It is important to highlight that we view this paper, and the lesson and worksheet that it presents, as a first step in the use of Santa Claus to support lessons on quantum teleportation.
This popular culture-based lesson is novel in the sense that it utilises the platform of Santa Claus, who most people would not traditionally connect to the concept of quantum teleportation.
Our main aim is to share this pedagogical approach with the community in the hope that others will recognise its uniqueness. 
In a further study, we plan to investigate the effectiveness of the lesson and worksheet in a classroom setting, and to collect student evaluations on the combination of the lesson and the worksheet.

Finally, this example presents a fun alternative to the classical Alice-Bob paradigm used to describe quantum communication protocols. 
While we do not seek to replace the successful Alice-Bob paradigm, we are confident that educators will recognise the novelty of our lesson plan, and implement it in the classroom when teaching the concept of quantum teleportation.

\acknowledgements
B.W.F and J.T. wish to thank the Heineken Prizes event for introducing them to each other during which B.W.F. suggested to J.T. the idea of the paper.
P.E. thanks Rebecca Kruse for useful feedback on an early version of this manuscript.
We thank Hans Norg for proof-reading the final manuscript. 
P.E. and J.T. acknowledge the support received by the Dutch National Growth Fund
(NGF), as part of the Quantum Delta NL programme. 
P.E. acknowledges the support received through the NWO-Quantum Technology
programme (Grant No. NGF.1623.23.006).
J.T. acknowledges the support received from the European Union’s Horizon Europe research and innovation programme through the ERC StG FINE-TEA-SQUAD (Grant No. 101040729). 
The views and opinions expressed here are solely
those of the authors and do not necessarily reflect those of the funding institutions. Neither
of the funding institutions can be held responsible for them.

\bibliography{references.bib}
\appendix
\onecolumngrid

\section{The full Santa-Claus story \label{app:story}}
It is Christmas, and this year, Santa Claus will be delivering one type of present to every child around the world - a book. 

Unfortunately, hiding presents in all homes across the world is a time-consuming task, which takes the whole year and cannot be completed in a single night.
Santa would love to plan ahead, and he would love to distribute the books well ahead of time.
However, he will only receive the children's present requests at the start of December. 
Santa and his elves, therefore, resort to quantum teleportation to deliver the books on time for Christmas Eve.

First, before any present book is delivered to a home, Santa and the elves use a special Christmas entanglement machine to entangle the present book with the workshop book at the workshop.
The two books are placed side-by-side in the machine, and the machine then decides if the content on any given page in both books will be either blank or full of text. 
Note, the entanglement machine cannot tell what the actual content is and nobody is allowed to open the books afterwards until the full protocol is completed.

After the entangling procedure, nobody is allowed to open the book and the machine is not allowed to `read' the text on a given page.
This would correspond to measuring and the superposition would vanish.
We would be left with purely classical and random information, and not with a quantum state.

The elves' job is not done after entangling the books.
One half of the entangled book-pair has to find its way to the children's homes.
Santa's elves are experts at hiding the present books in homes, and do so in a top-secret manner.
This arduous task is completed over the course of the year so that the elves have time to visit all homes around the world. 
The actual present, will only be teleported on Christmas Eve.

Of course, it can happen that a child finds the present book at home, which would ruin their Christmas present from Santa Claus. 
This is not an issue though because if a child were to open the present book without proper completion of the quantum teleportation process, the entangled state would be broken. 
As a result, the child would only see gibberish on the pages. 

The next stage of the process begins just before Santa Claus leaves his workshop on Christmas Eve.

First, for each child, the elves entangle the present book with the already entangled book at the workshop using the special Christmas entanglement machine. 
As before, the books are placed side-by-side in the entanglement machine, and the machine assesses whether on any given page from both books, there will be either a page full of text or a blank page in the two books. 
Then the elves add that information about the entanglement to a ledger for Santa Claus known as the \textit{Ledger of Good Nature} to keep a record of the entanglement process. 
Whether a child receives a book on Christmas or not is influenced by their level of good nature over the course of the year. 

Before Santa Claus is given the \textit{Ledge of Good Nature}, a ledger that contains a list of the books that each child are due to receive as presents on Christmas Eve with the book that they receive influenced in part by their level of good nature over the course of the past year. 

On Christmas Eve, Santa Claus travels around the world to deliver the presents to the children. 
In effect, he travels around the world to complete the quantum teleportation process. 

During the journey, Santa Claus checks the \textit{Ledge of Good Nature} to see whether the child will receive a present or not.
If the child has been well behaved during the year, then the child will get the book that they have asked for. 
However, if the ledger indicates that the child has misbehaved too much over the course of the year, the child will not get the book that they asked for. 
Instead, they will get random, gibberish text in their book. 

With this in mind, Santa Claus must complete the final step of the quantum teleportation by travelling around the world. 
When Santa Claus arrives at the home of each child, he quickly locates the hidden book, and corrects its contents according to the information in his ledger.
He sets up his present scanner with the right information from the \textit{Ledger of Good Nature}, which adapts the book's contents to ensure that the present book reads the same as the originally intended book.

On Christmas morning, when the child opens the book, they can open the book without worrying about seeing gibberish as a result of breaking the quantum teleportation process. 
Of course, if the child has been deemed to have misbehaved over the year, they will receive a book filled with gibberish text anyway!

Unfortunately, Santa Claus and the elves cannot use the books at the Arctic Circle or the books at the children's home again.
After all, it was a quantum teleportation and not a quantum copying protocol.
During the next year, the elves will have to entangle books again and redistribute them.
Added to that, they will have to print new books to be sent around.
The previous ones have been measured and no longer contain quantum information.

\section{Mathematical treatment of quantum teleportation\label{app:teleportation}}
This appendix is a more mathematical treatment of quantum teleportation.
It is not primarily aimed at the students, but rather to deepen the understanding of the educator.
The computation will be performed in two notations.
On the one hand, we use the bra-ket notation commonly used in quantum mechanics.
The symbol $\ket{\cdot}$ is called a ket and is the physicist's notation for a vector.
It is easier to perform the computations here, but the notation is rarely used in the classroom.
Thus, we translate all equations directly into matrix vector computations, which are equivalent, but usually more cumbersome to write.

\subsection{Notation and Background}
The goal of quantum teleportation is to transfer the quantum state $\ket{\psi}=\alpha\ket{0}+\beta\ket{1}$ from Alice to Bob.
Expressed as a vector, this equation reads
\begin{align}
    \ket{\psi}=\alpha\ket{0}+\beta\ket{1}=\alpha\mqty(1\\0)+\beta\mqty(0\\1)=\mqty(\alpha\\\beta).
\end{align}
Here, we used the definitions of $\ket{0}$ and $\ket{1}$ in Eq.~\eqref{eq:qubit_computational_basis}. 
In general, the coefficients can be complex, $\alpha,\beta\in\mathbb{C}$.
For the sake of simplicity, we will assume for the rest of the example that the coefficients are real $\alpha,\beta\in\mathbb{R}$.
The same computation works for complex coefficients.

As described in Section~\ref{sec:sub_entanglement}, Alice and Bob have a resource to perform the quantum teleportation.
An entangled state $\ket{\phi}$, where one of the qubits is located at Alice's side and the other is located at Bob's.
In the main text, we wrote the state as $\ket{\phi}=\frac{1}{\sqrt{2}}(\ket{0_A0_B}+\ket{1_A1_B})$.
This state consists of two qubits: one on Alice's side and one on Bob's side.
While a single qubit has two possible states, $\ket{0}$ and $\ket{1}$, this system as four possible states: $\ket{00}$, $\ket{10}$, $\ket{01}$ and $\ket{11}$.
The state $\ket{\phi}$ can also be written as a vector
\begin{align}
    \ket{\phi}=\frac{1}{\sqrt{2}}(\ket{0_A0_B}+\ket{1_A1_B})
    =\frac{1}{\sqrt{2}}\left(\mqty(1\\0\\0\\0)+\mqty(0\\0\\0\\1)\right)= \mqty(1/\sqrt{2}\\0\\0\\1/\sqrt{2})
    \label{eq:app_entangled_state_vector}
\end{align}

Instead of listing the basis states of a system, there is a more structured approach to combining quantum systems.
By taking the tensor product, denoted by $\otimes$, the vectors grow automatically to the appropriate size.
The tensor product takes two matrices of size $m \times n$ and $r\times k$ and returns a matrix of size $mr\times nk$.
A vector is interpreted as a $n\times 1$ matrix and the same rules apply.
We can directly apply the tensor product to two one-qubit vectors and see that we obtain the same vectors as we had before in Eq.~\eqref{eq:app_entangled_state_vector}
\begin{align}
    \ket{01}&=\ket{0}\otimes\ket{1}=\mqty(1\\0)\otimes\mqty(0\\1) =\mqty(1\mqty(0\\1)\\0\mqty(0\\1))=\mqty(0\\1\\0\\0)\\
    \ket{00}&=\ket{0}\otimes\ket{0}=\mqty(1\\0)\otimes\mqty(1\\0) =\mqty(1\mqty(1\\0)\\0\mqty(1\\0))=\mqty(1\\0\\0\\0)
\end{align}
The notation of multiple numbers in the same ket, as in $\ket{01}$, implicitly expresses a tensor product: $\ket{01}=\ket{0}\otimes\ket{1}$.

Using the same procedure, we notice that a three-qubit system must have eight-dimensional vectors.
When we first combine two single qubit systems, we obtain a four-dimensional object like $\ket{00}$.
Combining it with another two-dimensional qubit, we obtain an eight-dimensional state like $\ket{000}$.

For the protocol we do not need only qubits, but also operators acting on the qubits.
Operators are manipulations that change the state of the qubit.
We start with an example on one qubit.
The operation of flipping a qubit from $\ket{0}$ to $\ket{1}$ is performed by the $X$ operator.
You can think of the $X$ operator as the logical $\mathrm{NOT}$.
In matrix language, we express the operator as
\begin{align}
    X=\mqty(0&1\\1&0).
\end{align}
We can check directly that it acts in the proper way
\begin{align}
    X\ket{0}=\mqty(0&1\\1&0)\mqty(1\\0)=\mqty(0\\1);\quad X\ket{1}=\mqty(0&1\\1&0)\mqty(0\\1)=\mqty(1\\0).
\end{align}
The state $\ket{0}$ is transformed into $\ket{1}$, and vice versa.

The quantum teleportation protocol needs only two more operators.
The first one is the $Z$ operator and it leaves the computational basis states $\ket{0}$ and $\ket{1}$ invariant.
It only adds a minus sign to $\ket{1}$.
Similarly, we apply the $Z$ gate as
\begin{align}
    Z=\mqty(\dmat{1,-1});\quad
    Z\ket{0}=\mqty(\dmat{1,-1})\mqty(1\\0)=\mqty(1\\0);\quad Z\ket{1}=\mqty(\dmat{1,-1})\mqty(0\\1)=-\mqty(0\\1).
\end{align}
In addition to the $X$ and the $Z$ gate, we need the so-called Hadamard gate.
It helps us to generate a superposition from the states $\ket{0}$ and $\ket{1}$.
We can write it as
\begin{align}
    H=\frac {1}{\sqrt {2}}\mqty(1&1\\1&-1).
\end{align}
When it acts on $\ket{0}$ and $\ket{1}$, we obtain
\begin{align}
    H\ket{0}&= \frac{1}{\sqrt{2}}\mqty(1&1\\1&-1)\mqty(1\\0)=\frac{1}{\sqrt{2}}\mqty(1\\1)=\frac{1}{\sqrt{2}}(\ket{0}+\ket{1}),\\
    H\ket{1}&= \frac{1}{\sqrt{2}}\mqty(1&1\\1&-1)\mqty(0\\1)=\frac{1}{\sqrt{2}}\mqty(1\\-1)=\frac{1}{\sqrt{2}}(\ket{0}-\ket{1}).
\end{align}

Acting on a single qubit alone will not be enough as Figure~\ref{fig:teleportation} already suggests with the bubbles wrapping around two qubits.
We need one two qubit gate for the protocol: the so-called controlled-not $\mathrm{cNOT}$ operation. 
Since it acts on two qubits, it must be a $4\times 4$ matrix instead of a $2\times 2$ matrix: two-qubit states have four dimensions.
The idea of the $\mathrm{cNOT}$ is to flip the second qubit depending on the first qubit (called control qubit).
If the first qubit is in state $\ket{1}$, the second qubit will be flipped.
The matrix is given by
\begin{align}
    cNOT=\mqty(1 & 0 & 0 & 0\\
               0 & 1 & 0 & 0\\
               0 & 0 & 0 & 1\\
               0 & 0 & 1 & 0).
\end{align}
The action of the $\mathrm{cNOT}$ is illustrated by
\begin{align}
cNOT\ket{00}=\mqty(1 & 0 & 0 & 0\\
               0 & 1 & 0 & 0\\
               0 & 0 & 0 & 1\\
               0 & 0 & 1 & 0)\mqty(1\\0\\0\\0)=\mqty(1\\0\\0\\0)=\ket{00};\quad
cNOT\ket{10}=\mqty(1 & 0 & 0 & 0\\
               0 & 1 & 0 & 0\\
               0 & 0 & 0 & 1\\
               0 & 0 & 1 & 0)\mqty(0\\0\\1\\0)=\mqty(0\\0\\0\\1)=\ket{11}.
\end{align}
Note that the state of the first qubit always stays unchanged and the second qubit flips depending on the state of the first qubit.

\subsection{Quantum Teleportation}
Equipped with these tools, we are ready to write down the state of the full system and perform a quantum teleportation.
We write the state of the system including all three qubits as
\begin{align}
    \ket{S}&=\ket{\psi}\otimes\ket{\phi}=\left(\alpha\ket{0}+\beta\ket{1}\right)\otimes\frac{1}{\sqrt{2}}\left(\ket{00}+\ket{11}\right)\\
    &=\frac{1}{\sqrt{2}}\left(\alpha\ket{000}+\alpha\ket{011}+\beta\ket{100}+\beta\ket{111}\right).
    \label{app:eq_teleport_system}
\end{align}
Here, we used the distributive property of the tensor product:
we can directly compute
\begin{align}
    \left(\ket{0}+\ket{1}\right)\otimes\ket{1}=\ket{0}\otimes\ket{1}+\ket{1}\otimes\ket{1}=\ket{01}+\ket{11}
\end{align}
without writing the vector notation.
The numbers in the ket expressions like $\ket{011}$ follow the convention $\ket{A_1 A_2 B}$.
The first qubit $A_1$ on Alice's side holds the information that we would like to teleport.
Alice's second qubit $A_2$ is part of the Bell pair and the third qubit $B$ is Bob's qubit, also part of the entangled pair $A_2 B$.

\subsection{Operations on Alice's qubits}
The first step of the protocol is an entanglement operation between Alice's two qubits.
This operation is performed with a $\mathrm{cNOT}$ operation on both of Alice's qubits.
Note that all operation of the protocol are local.
That means that there are not two-qubit operations that act on a qubit of Alice and a qubit of Bob at the same time.
We obtain the state after the first operation $\ket{S_1}$ by applying the $\mathrm{cNOT}$ gate.
\begin{align}
    \ket{S_1}&=(\mathrm{cNOT}_{A_1A_2}\otimes \mathds{1}_B)\ket{S}\\
    &=\left(\mathrm{cNOT}_{A_1A_2}\otimes \mathds{1}_B\right)\frac{1}{\sqrt{2}}(\alpha\ket{000}+\alpha\ket{011}+\beta\ket{100}+\beta\ket{111})\\
    &=\frac{1}{\sqrt{2}}\left(\alpha\ket{000}+\alpha\ket{011}+\beta\ket{110}+\beta\ket{101}\right).
    \label{app:eq_teleport_step_1}
\end{align}
Before we write the same calculation in terms of matrices, let us have a look at the operator $\mathrm{cNOT}_{A_1A_2}\otimes \mathds{1}$.
As introduced above, $\mathrm{cNOT}$ acts on $2$ qubits, but $\ket{S}$ is a three qubit state.
The $\mathrm{cNOT}$ acts on the first two (Alice's) qubits, thus we have to define what happens on with the third qubit. 
since we do not want any action on the third qubit, we build the tensor product with the identity of size 2, written as $\mathds{1}$.

Now, we will do the same computation in Eq.~\eqref{app:eq_teleport_step_1} in terms of matrices.
This computation adds nothing new to the derivation, it is just a different of writing it.
First, we have to write down the matrix of the operation $\left(\mathrm{cNOT}_{A_1A_2}\otimes \mathds{1}_B\right)$
\begin{align}
\mathrm{cNOT}_{A_1A_2}\otimes \mathds{1}_B =
    \mqty(1 & 0 & 0 & 0\\
          0 & 1 & 0 & 0\\
          0 & 0 & 0 & 1\\
          0 & 0 & 1 & 0)  \otimes \mqty(\dmat{1,1})=\mqty( 1 & 0 & 0 & 0 & 0 & 0 & 0 & 0 \\
 0 & 1 & 0 & 0 & 0 & 0 & 0 & 0 \\
 0 & 0 & 1 & 0 & 0 & 0 & 0 & 0 \\
 0 & 0 & 0 & 1 & 0 & 0 & 0 & 0 \\
 0 & 0 & 0 & 0 & 0 & 0 & 1 & 0 \\
 0 & 0 & 0 & 0 & 0 & 0 & 0 & 1 \\
 0 & 0 & 0 & 0 & 1 & 0 & 0 & 0 \\
 0 & 0 & 0 & 0 & 0 & 1 & 0 & 0 \\).
\end{align}
After writing the state $\ket{S}$ [cf. Eq.~\eqref{app:eq_teleport_system}] as 
\begin{align}
    \ket{S}&=\ket{\psi}\otimes\ket{\phi}=\left(\alpha\ket{0}+\beta\ket{1}\right)\otimes\frac{1}{\sqrt{2}}\left(\ket{00}+\ket{11}\right)\\
    &=\mqty(\alpha\\\beta)\otimes \mqty(\frac{1}{\sqrt{2}}\\0\\0\\\frac{1}{\sqrt{2}})
    =\frac{1}{\sqrt{2}}\mqty(\alpha & 0 & 0 & \alpha & \beta & 0 & 0 & \beta)^T.
    \label{app:eq_state_system_vector_notation}
\end{align}
we can transform the operation in Eq.~\eqref{app:eq_teleport_step_1}
\begin{align}
    \ket{S_1}&=(\mathrm{cNOT}_{A_1A_2}\otimes \mathds{1}_B)\ket{S}
    =\frac{1}{\sqrt{2}}\mqty( 1 & 0 & 0 & 0 & 0 & 0 & 0 & 0 \\
                               0 & 1 & 0 & 0 & 0 & 0 & 0 & 0 \\
                               0 & 0 & 1 & 0 & 0 & 0 & 0 & 0 \\
                               0 & 0 & 0 & 1 & 0 & 0 & 0 & 0 \\
                               0 & 0 & 0 & 0 & 0 & 0 & 1 & 0 \\
                               0 & 0 & 0 & 0 & 0 & 0 & 0 & 1 \\
                               0 & 0 & 0 & 0 & 1 & 0 & 0 & 0 \\
                               0 & 0 & 0 & 0 & 0 & 1 & 0 & 0 \\)\mqty(\alpha \\ 0 \\ 0 \\ \alpha \\ \beta \\ 0 \\ 0 \\ \beta)\\
    &=\frac{1}{\sqrt{2}} \mqty(\alpha  & 0 & 0 & \alpha & 0 & \beta & \beta &0 )^T.
\end{align}
Here, we wrote the last vector in its transposed form for convenience.
At this point we will continue with the algorithm in the bra-ket notation and show the equivalence with matrix calculus again after the next step.
To see the direct correspondence between the last line of Eq.~\eqref{app:eq_teleport_system} and Eq.~\eqref{app:eq_state_system_vector_notation}, we give the full basis for three qubits in Appendix~\ref{app:basis_three_qubits}.

The second operation of the quantum teleportation protocol is a Hadamard gate acting on the first qubit $A_1$.
Note that this is still the first step as described in the main text.
\begin{align}
    \ket{S_2}&=(H_{A_1}\otimes \mathds{1}\otimes\mathds{1})\ket{S_1}
    =\frac{1}{\sqrt{2}}H_1\left(\alpha\ket{000}+\alpha\ket{011}+\beta\ket{110}+\beta\ket{101}\right)\\
    =&\frac{1}{2}\left(\alpha\left[\ket{000}+\ket{100}+\ket{011}+\ket{111}\right]+\beta\left[\ket{010}+\ket{001}-\ket{110}-\ket{101}\right]\right)
    \label{app:eq_teleport_step_2}
\end{align}
We can perform the equivalent operation by first writing the Hadamard gate in the three qubit basis as
\begin{align}
    H_{A_1}\otimes\mathds{1}\otimes\mathds{1}= \mqty(
     \frac{1}{\sqrt{2}} & 0 & 0 & 0 & \frac{1}{\sqrt{2}} & 0 & 0 & 0
   \\
 0 & \frac{1}{\sqrt{2}} & 0 & 0 & 0 & \frac{1}{\sqrt{2}} & 0 & 0
   \\
 0 & 0 & \frac{1}{\sqrt{2}} & 0 & 0 & 0 & \frac{1}{\sqrt{2}} & 0
   \\
 0 & 0 & 0 & \frac{1}{\sqrt{2}} & 0 & 0 & 0 & \frac{1}{\sqrt{2}}
   \\
 \frac{1}{\sqrt{2}} & 0 & 0 & 0 & -\frac{1}{\sqrt{2}} & 0 & 0 & 0
   \\
 0 & \frac{1}{\sqrt{2}} & 0 & 0 & 0 & -\frac{1}{\sqrt{2}} & 0 & 0
   \\
 0 & 0 & \frac{1}{\sqrt{2}} & 0 & 0 & 0 & -\frac{1}{\sqrt{2}} & 0
   \\
 0 & 0 & 0 & \frac{1}{\sqrt{2}} & 0 & 0 & 0 & -\frac{1}{\sqrt{2}}).
\end{align}
The computation in Eq.~\eqref{app:eq_teleport_step_2} is equivalent to
\begin{align}
    \ket{S_2}&=(H_{A_1}\otimes \mathds{1}\otimes\mathds{1})\ket{S_1}\\
    &=\frac{1}{\sqrt{2}}\mqty(\frac{1}{\sqrt{2}} & 0 & 0 & 0 & \frac{1}{\sqrt{2}} & 0 & 0 & 0
   \\
 0 & \frac{1}{\sqrt{2}} & 0 & 0 & 0 & \frac{1}{\sqrt{2}} & 0 & 0
   \\
 0 & 0 & \frac{1}{\sqrt{2}} & 0 & 0 & 0 & \frac{1}{\sqrt{2}} & 0
   \\
 0 & 0 & 0 & \frac{1}{\sqrt{2}} & 0 & 0 & 0 & \frac{1}{\sqrt{2}}
   \\
 \frac{1}{\sqrt{2}} & 0 & 0 & 0 & -\frac{1}{\sqrt{2}} & 0 & 0 & 0
   \\
 0 & \frac{1}{\sqrt{2}} & 0 & 0 & 0 & -\frac{1}{\sqrt{2}} & 0 & 0
   \\
 0 & 0 & \frac{1}{\sqrt{2}} & 0 & 0 & 0 & -\frac{1}{\sqrt{2}} & 0
   \\
 0 & 0 & 0 & \frac{1}{\sqrt{2}} & 0 & 0 & 0 & -\frac{1}{\sqrt{2}})  \mqty(\alpha  \\ 0 \\ 0 \\ \alpha \\ 0 \\ \beta \\ \beta \\0 )\\
 &= \frac{1}{2} \mqty(\alpha & \beta & \beta & \alpha & \alpha & -\beta & -\beta & \alpha )^T
\end{align}

\subsection{Measurement on Alice's side}
Starting from Eq.~\eqref{app:eq_teleport_step_2}, we can use the distributive property of the tensor product in the opposite direction and write equivalently
\begin{align}
    \begin{aligned}
        \ket{S_2}=&\frac{1}{2}\left(\ket{00}\otimes(\alpha\ket{0}+\beta\ket{1})\right)+\\
                  &\frac{1}{2}\left(\ket{01}\otimes(\alpha\ket{1}+\beta\ket{0})\right)+\\
                  &\frac{1}{2}\left(\ket{10}\otimes(\alpha\ket{0}-\beta\ket{1})\right)+\\
                  &\frac{1}{2}\left(\ket{11}\otimes(\alpha\ket{1}-\beta\ket{0})\right)
  \end{aligned}
  \label{app:eq_superposition_alice_bob}
\end{align}
On the first glance this might look like a mathematical trick, but it actually has a physical interpretation.
The coefficients $\alpha$ and $\beta$ are now prefactors of the third qubit.
This is Bob's qubit!
That means that the information about the original state on Alice's side $\ket{\psi}=\alpha\ket{0}+\beta\ket{1}$ is now on Bob's side.
The same step is much harder to see in matrix notation, but the decomposition is equally possible in the vector-matrix picture.

One problem remains: the prefactors are not precisely correct in all cases.
Sometimes, they are in front of the wrong qubit ($\alpha\ket{1}$ instead of $\alpha\ket{0}$) and sometimes the signs do not match ($-\beta$ instead of $\beta$).
We can fix this by allowing Alice to measure her qubits in the computational basis.
By measuring, she obtains the bit values of her states, i.e. she reads out $01$ for the state $\ket{01}$.
With this additional information, the problem is solved!
The superposition in Eq.~\eqref{app:eq_superposition_alice_bob} collapses and only one part of it is realized.
Since Alice knows her measurement, e.g. $01$, she knows that Bob's qubit must be in the state $\frac{1}{\sqrt{2}}\left(\alpha\ket{1}+\beta\ket{0}\right)$.

\subsection{Correction of Bob's qubit}
Alice submits her bit string to Bob and he knows now the operation that he has to perform to obtain the proper state $\ket{\psi}$.
If Alice measures, for example, $01$, he knows that the signs of the coefficients are correct, but they are in front of the wrong states.
By applying an $X$ gate to this state, he will obtain the correct state
\begin{align}
    X\left(\beta\ket{0}+\alpha\ket{1}\right)=\alpha\ket{0}+\beta\ket{1}.
\end{align}
All four possible outcomes and the corresponding operations are listed in Table~\ref{tab:compensation}.
\begin{table}
    \centering
    \begin{tabular}{ccc}
        \toprule
        Bob's state & Alice's measuremnt & Bob's operation\\
        \midrule
        $\alpha\ket{0}+\beta\ket{1}$ & $00$ & $\mathds{1}$\\
        $\alpha\ket{1}+\beta\ket{0}$ & $01$ & $X$\\
        $\alpha\ket{0}-\beta\ket{1}$ & $10$ & $Z$\\
        $\alpha\ket{1}-\beta\ket{0}$ & $11$ & $ZX$\\
        \bottomrule
    \end{tabular}
    \caption{Operations on Bob's side to adapt the state according to Alice's measurement}
    \label{tab:compensation}
\end{table}

\subsection{Additional Remarks}
The measurement on Alice's side is not only essential for the protocol, it is also important to avoid a contradiction with the no-cloning theorem~\cite{park_concept_1970,wootters_single_1982}.
The no-cloning theorem is a so-called no-go theorem, a proof that shows the impossibility of something.
In this case, the no-cloning theorem states that it is impossible to copy an arbitrary quantum state.
Since we do not know anything about the quantum state $\ket{\psi}$ of Alice, i.e. $\alpha$ and $\beta$ are unkown, we should not be allowed to copy it.
And indeed, we are not copying it.
Alice destroys her quantum state by measuring her two qubits. 
Thus, the state is transferred (or teleported) and not copied.

The second remark is related to special relativity.
One of the postulates of special relativity is that no information can be transferred faster than light.
If we could transfer the state from Alice to Bob without Alice telling Bob about her measurements, we would have indeed transferred information faster than light.
However, as Bob will not know whether he has the correct state or a modified one until Alice tells him her measurements.
Alice will communicate her measurement results to Bob via a classical communication channel, e.g. telephone, email or any other means of non-quantum communication.
This communication is limited by the speed of light.
In the end, no information is transmitted faster than light and quantum teleportation does not contradict special relativity.

\subsection{Full basis for three qubits\label{app:basis_three_qubits}}
As a reference and to ease the translation between bra-ket notation and matrix-vector notation, we provide here the all computational basis states for three qubits.
\begin{align}
    \begin{aligned}
        \ket{000}&=\mqty(1&0&0&0&0&0&0&0)^T\\
        \ket{001}&=\mqty(0&1&0&0&0&0&0&0)^T\\
        \ket{010}&=\mqty(0&0&1&0&0&0&0&0)^T\\
        \ket{011}&=\mqty(0&0&0&1&0&0&0&0)^T\\
        \ket{100}&=\mqty(0&0&0&0&1&0&0&0)^T\\
        \ket{101}&=\mqty(0&0&0&0&0&1&0&0)^T\\
        \ket{110}&=\mqty(0&0&0&0&0&0&1&0)^T\\
        \ket{111}&=\mqty(0&0&0&0&0&0&0&1)^T
    \end{aligned}
    \label{app:eq_basis_three_qubits}
\end{align}

\section{Santa Claus worksheet\label{app:worksheet}}
To make the worksheet as easy to use as possible, we included it as a full page on the following page.
\clearpage
\newpage
\pagestyle{empty}
\begin{center}
\includegraphics[width=\textwidth]{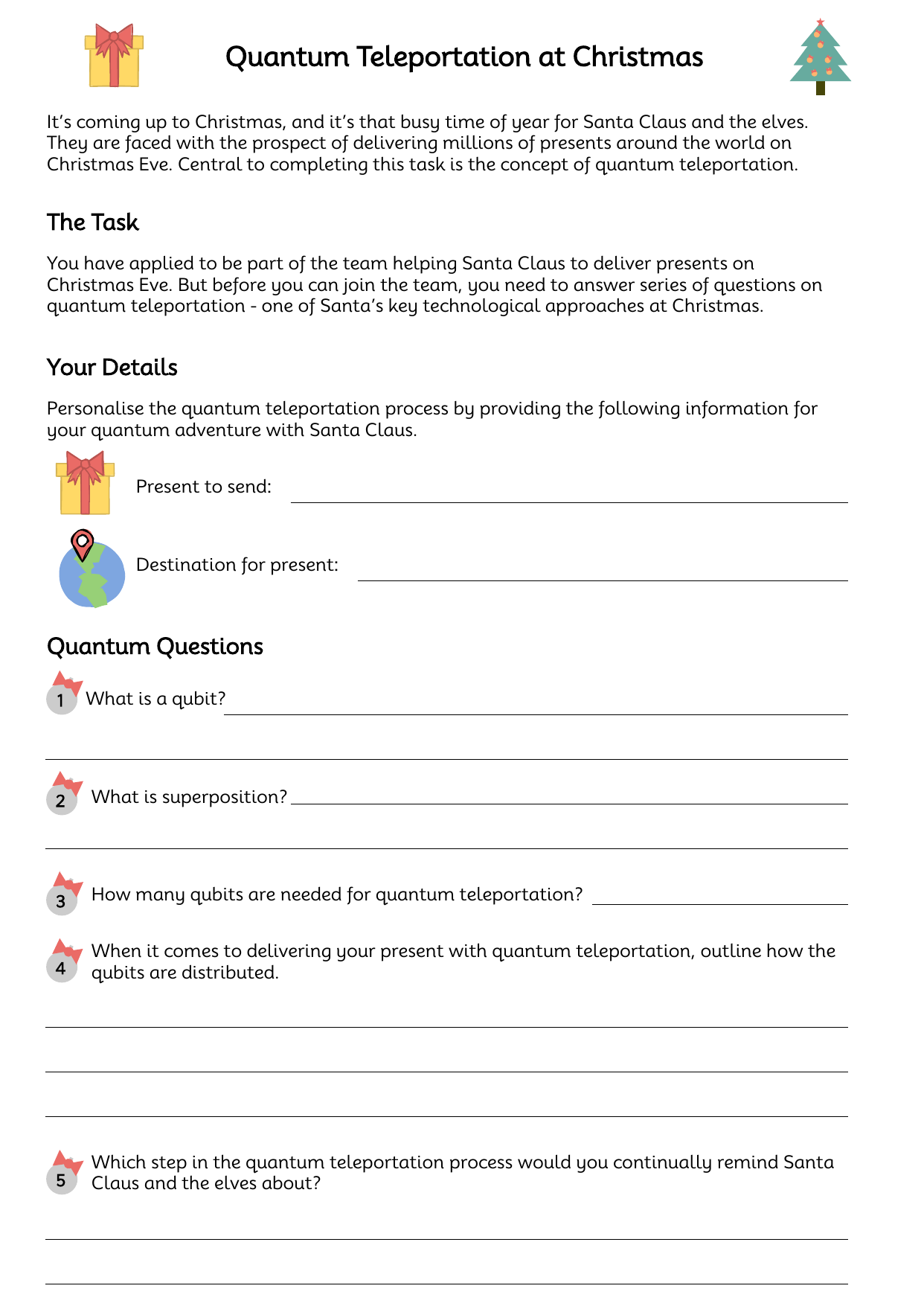}
\end{center}

\end{document}